\documentclass[12pt]{iopart}
\usepackage{iopams,amsfonts,amssymb,amsthm}
\usepackage{graphicx}
\usepackage{rawfonts}
\usepackage{epstopdf}
\usepackage[square,sort&compress,numbers]{natbib}
\newcommand\newblock{\hskip .11em\@plus.33em\@minus.07em}

\usepackage{color}

\begin{document}

\title{Knot probabilities in equilateral random polygons}
\author{A Xiong$^1$, A J Taylor$^2$, M R Dennis$^{1,2,3}$ and S G  Whittington$^4$ }
\address{
{}$^1$School of Physics and Astronomy, University of Birmingham, Birmingham B15 2TT, UK \\
{}$^2$H H Wills Physics Laboratory, University of Bristol, Bristol BS8 1TL, UK \\
{}$^3$EPSRC Centre for Doctoral Training in Topological Design, University of Birmingham, Birmingham B15 2TT, UK \\
{}$^4$Department of Chemistry, University of Toronto, Toronto, M5S 3H6, Canada.

}

\begin{abstract}
We consider the probability of knotting in equilateral random polygons in Euclidean 3-dimensional space, which model, for instance, random polymers.
Results from an extensive Monte Carlo dataset of random polygons indicate a universal scaling formula for the knotting probability with the number of edges.
This scaling formula involves an exponential function, independent of knot type, with a power law factor that depends on the number of prime components of the knot.
The unknot, appearing as a composite knot with zero components, scales with a small negative power law, contrasting with previous studies that indicated a purely exponential scaling.
The methodology incorporates several improvements over previous investigations: our random polygon data set is generated using a fast, unbiased algorithm, and knotting is detected using an optimised set of knot invariants based on the Alexander polynomial.
\end{abstract}

\section{Introduction}

\begin{figure}
\includegraphics[width=\textwidth]{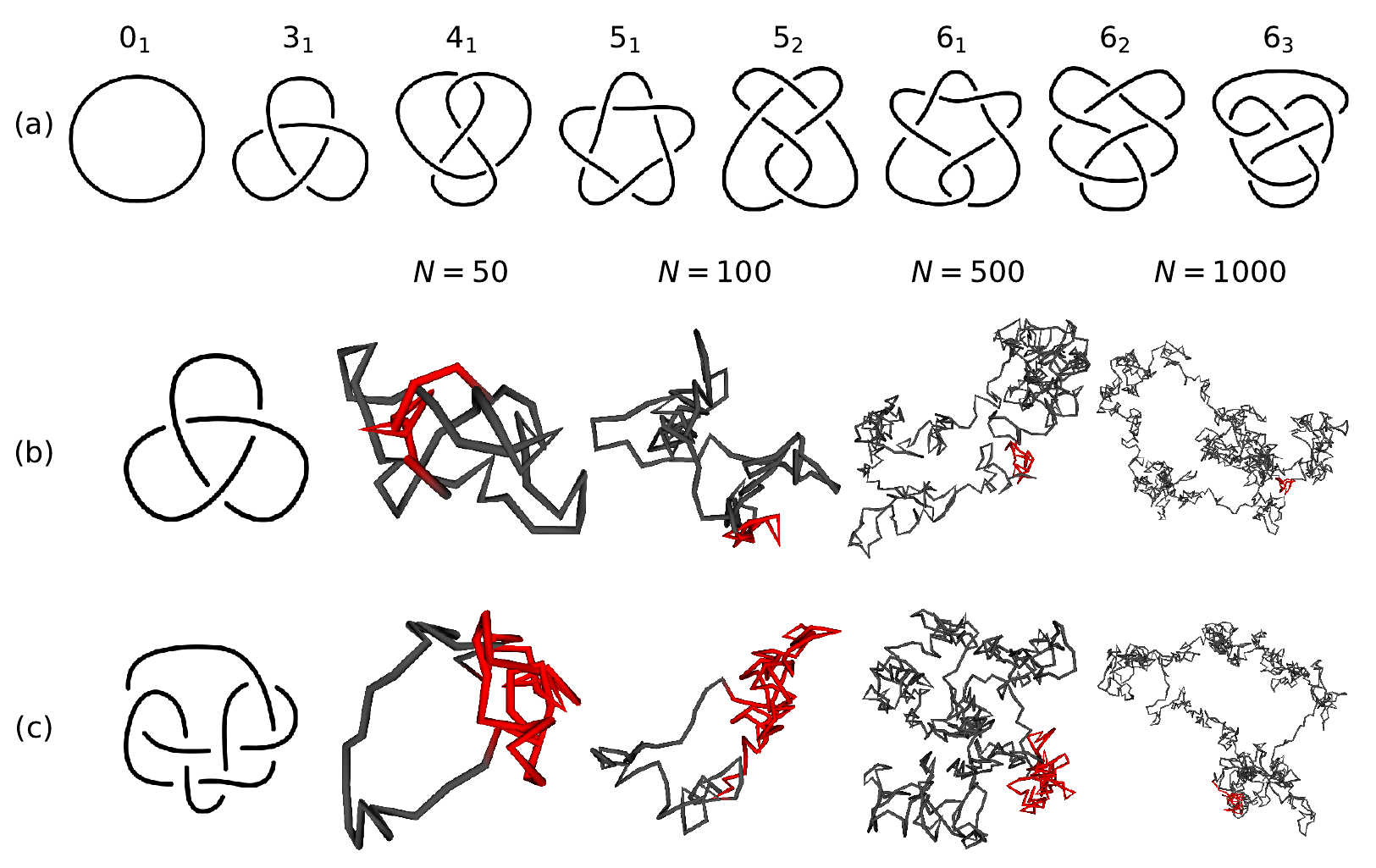} 
  \caption{Some simple knots, manifested in closed random walks of different lengths. 
    (a) shows the unknot $0_1$ and the seven different prime knots with minimum crossing number $n_{\mathrm{c}} \le 6$. 
    The knots (b) $3_1$ and (c) $7_6$ are realised in equilateral random polygons of different lengths, up to $1000$ edges. 
    For each polygon, the knotted region is highlighted in red.}
  \label{fig:knots_and_random_walks}
\end{figure}

The tendency of long random filaments to become knotted is familiar to everyone carrying headphone cables in their pocket.  
It seems natural to expect that the probability that a random closed curve in three dimensions is knotted increases with its length.  
Random knotting---especially in closed random walks---has been studied at least since the 1960s.
It was conjectured~\citep{frisch61,delbruck62} that sufficiently long linear polymers in dilute solution, undergoing a ring closure reaction, would produce knotted ring polymers with high probability.
The study of knotted random walks has been associated with knotted polymers ever since, employing insight and techniques from geometry, topology and statistical mechanics.
Analytical results are rare \cite{diao94,diao95}, so the problem is most naturally studied with computers. 
Closed random walks of sufficient length must be generated, whose knotting is analysed to investigate the asymptotics.
Topologically distinct kinds of knot are classified (the simplest examples are shown in Figure \ref{fig:knots_and_random_walks}(a)), and so we can ask what the probabilities of different \emph{knot types} are in closed random walks.

A natural statistical model for this investigation, extensively studied, is the ensemble of \emph{equilateral random polygons} \cite{diao95}.  
These are piecewise linear embeddings of the circle in $\mathbb{R}^3$ such that each edge has unit length: they are effectively random walks in three-dimensional Euclidean space, conditioned to return to their starting point.
(The equilateral condition can be relaxed to study walks whose edge lengths have some other distribution~\cite{diao94}.)
Examples of random polygons of different edge numbers with two specific knot types are shown in Figure \ref{fig:knots_and_random_walks} (b) and (c).
\citet{diao95} has demonstrated that the probability of an equilateral random polygon being \emph{unknotted} tends to zero as the number of edges increases. 
Many studies of knotted random walks have followed, generating different statistical ensembles, utilising larger datasets as numerical power has improved, and with increasingly sophisticated knot type analyses~\citep{millett2005,deguchi94,tsurusaki95,deguchi97,matsuda03,uehara15,uehara17}.
Several facts agreeing with large-$N$ asymptotics are supported, such as the probability of composite knots increasing with $N$ \cite{diao95}.

Here we revisit these questions, bringing contemporary large-scale computing resources to bear on the problem by generating many millions of random polygons by a Monte Carlo routine.
Our methodology incorporates several improvements over previous studies, including:
\begin{itemize}
\item large datasets, targeting $10^7$ random polygons with edge number $N$ up to $4000$ steps (compared to $\sim 2 \times 10^6$ from other sources), though much of our analysis is based on polygons with $N \le 3000$ where the statistics are better;
\item utilising the algorithm of \citet{cantarella16}, which samples equilateral random polygons correctly and quickly; 
\item identifying knots using a set of optimised numerical knot invariants.
\end{itemize}

As we shall describe, our results are consistent with the probability $P_K(N)$ of a particular knot of type $K$ occurring in a random polygon with $N$ sides, having the form
\begin{equation}
  \label{eq:ansatz}
  P_K(N) = C_K N^{v_K} \exp \left( - \frac{N}{N_K} \right) \left[ 1 + \beta_K N^{-\Delta} + \gamma_K N^{-1} + o(N^{-1})\right].
\end{equation}
This expression incorporates an overall exponential decay, with decay parameter $N_K$, combined with a power law exponent $v_K$ and other asymptotic corrections to scaling ($\beta_K$, $\gamma_K$ and $\Delta$).
The first three factors involving constants $C_K$, $v_K$ and $N_K$ are similar to the analogous knotting probability for self-avoiding polygons of $N$ edges on a cubic lattice~\cite{orlandini98}, which is well-grounded in a range of different random polygon models~\cite{deguchi94,tsurusaki95,deguchi97,uehara15,uehara17}. 
The small-scale correction terms to scaling are less well studied, and the term in square brackets in (\ref{eq:ansatz}) is guessed based on the behaviour of lattice models~\cite{orlandini98}.
The value of the confluent correction exponent $\Delta$ is not known precisely, but we assume that $\Delta = 1/2$ \cite{orlandini98}.

The form (\ref{eq:ansatz}) is supported by our observations of many knot types occurring for both smaller and larger $N$, including prime knots with up to nine crossings and composite knots with up to five components, and we present data for eight-crossing prime knots and four-component composites. 
Knot terminology is explained in Section~\ref{sec:methods_topological_identification}.
Our results suggest that the exponential parameter $N_K$ is \emph{universal for all knot types}, $N_K = N_0$, both prime and composite \cite{deguchi94}.

From our data, the power law exponent appears to be $v_K = v_0 + n_{\mathrm{p}}(K)$, with universal constant $v_0$ and $n_{\mathrm{p}}(K)$, \emph{the number of prime components} of $K$.
Prime knots have one prime component ($n_{\mathrm{p}} = 1$), composite knots have more than one prime component ($n_{\mathrm{p}} > 1$).
This form of exponent is supported by an argument that the knotted components are on average localised and relatively small along the curve~\cite{katritch00} and unentangled with one another~\cite{tsurusaki95}.  
If we had a pattern theorem for the unknot, then, under these assumptions,  the composite exponent would be the sum of the exponent for the unknot and the number of prime components in the knot decomposition. 
This behaviour is comparable to lattice models, for which there is evidence that $v_K = n_{\mathrm{p}}$~\cite{orlandini96,orlandini98,baiesi2010}, up to small corrections.
In this sense, $v_K$ controls the asymptotic \emph{relative} frequency of composite knots with different $n_{\mathrm{p}}$: the knot with more components will always eventually be more common. 
Knotted random polygons appear to follow this behaviour, with a small negative offset $v_0$.
This deviation of $v_K$ from $n_{\mathrm{p}}$ has been seen elsewhere~\cite{uehara17}, so the deviation from the lattice result seems to be typical of unconstrained random polygons.

Notably, the unknot appears as the composite knot with zero prime components, $n_{\mathrm{p}} = 0$, with the small offset and the same exponential parameter $N_0$.
That the unknot scales as a ``zero component knot'', rather than with no power law and possibly a different $N_K$, is a new observation from our data.

The amplitude $C_K$ depends on knot type and is the only parameter that differentiates between prime knots or composite knots with the same $n_{\mathrm{p}}$ up to corrections to scaling.

From the best fits to our data, the values of these universal constants are $N_0 = 259.3 \pm 0.2$ for equilateral random polygons, and $v_0 = -0.190 \pm 0.001$. 
Since $-1/2 < v_0 < 0$, the unknot acquires a \emph{negative power law scaling} in addition to the well-established exponential decay with $N$.
This contrasts with previous studies, where the unknot probability was interpreted as scaling exponentially with no power law.
Our results are not necessarily incompatible with prior investigations, in which the errors (due to smaller samples) are larger.

We also made preliminary investigations for an ensemble of non-equilateral random polygons. We used an ensemble of closed random polygons based on quaternions introduced in \cite{cantarella13}, which we call the \emph{quaternionic model}. 
This has the advantage of being very fast and straightforward to implement numerically, and the quaternionic polygons have edgelengths sampled from beta distributions. 
For the quaternionic model we find the same power law, $v_0 = -0.19 \pm 0.03$, and a different exponent, $N_K = 430.5 \pm 1$.
The similarity of $v_0$ and difference of $N_0$ is consistent with expectations of knot scaling. 
  We will not describe many features of this model, but the broad findings are consistent with the random polygons.

This type of numerical analysis is fundamentally limited: longer random polygons are not only more computationally expensive to analyse, but may adopt a vast plethora of prime and composite knot types. 
The chance of a specific knot type occurring for large $N$ therefore drops dramatically.
Furthermore, more complex knot types are harder to identify numerically, and it becomes hard to find topological invariants that robustly distinguish them in realistic timescales.  
We make some simple estimates of the misidentification rate to support our main conclusions, but such difficulties limit the maximum $N$ for which reliable data can be found.

Furthermore, given that $N_K = N_0$ and $v_K = v_0 + 1 > 0$ for all prime knots, the Ansatz (\ref{eq:ansatz}) suggests that all prime knots have a maximum probability at $N \approx N_0 (1+v_0)$ (with error depending on corrections to scaling).
Clearly, knots with a large number of crossings (of $N_0 (v_0+1)$ crossings or more) cannot have this maximum, and indeed we show that the position of the maximum drifts, depending on the correction to scaling parameters $\beta_K$ and $\gamma_K$.
Nevertheless, (\ref{eq:ansatz}) gives a good agreement with the data of a significant number of the commonest random knots, both prime, composite and the unknot.

The outline of the remainder of this paper is as follows.
The various subsections of Section \ref{sec:methods} provide the details of random walk generation, knot detection and classification, and numerical parameter choices. 
In Section~\ref{sec:results} we describe our results, and we conclude with a brief discussion in Section~\ref{sec:discussion}. 
In Section 4 we summarize our results from the quaternionic random walk model.
Before this, however, we briefly summarise the knotting properties of random polygons confined to lattices, justifying the form of the Ansatz (\ref{eq:ansatz}).

\subsection{Knotting in lattice polygons}
\label{sec:lattice_polygons}

Although there are few analytic results \cite{diao95}  to test against the numerical results for the random polygons, some rigorous results are available for random \emph{lattice polygons} (simple closed curves embedded in a three-dimensional lattice such as the simple cubic lattice, $\mathbb{Z}^3$).
These rigorous results \cite{sumners88,pippenger89} guide our questions about the behaviour of random polygons in the continuum.

Writing $p_N$ for the number of polygons in the simple cubic lattice with $N$ edges, up to translation, clearly we have $p_N=0$ if $N$ is odd, $p_4=3$ and $p_6=22$.
Hammersley \cite{hammersley61} showed that the limit, taken through even values of $N$,
\begin{equation}
  \lim_{N\to\infty} N^{-1} \log p_N \equiv \log \mu~, 
\end{equation}
exists and the \emph{growth constant} $\mu$ satisfies $3 < \mu < 5$.  
If $p_N(\emptyset) \equiv p_N^0$ is the number of $N$-edge polygons that are unknotted, then \cite{sumners88,pippenger89}
\begin{equation}
   \lim_{N\to\infty} N^{-1} \log p_N^0 \equiv \log \mu_0,
\end{equation}
and $\mu_0 < \mu$, i.e.~unknotted polygons are exponentially rare in the set of lattice polygons.  
If $p_N(K)$ denotes the number of $N$-edge polygons of knot type $K$ then, similarly,
\begin{equation}
   \log \mu_0 \le \liminf_{N\to\infty} N^{-1} \log p_N(K) \le \limsup_{N\to\infty} N^{-1} \log p_N(K) < \log \mu~,
\end{equation}
so polygons with any fixed knot type are also exponentially rare. 
The existence of the limit has not been proved for any knot type other than the unknot, and it has not been proved whether or not the exponential growth rate is independent of knot type.

Although these rigorous results give interesting information about knot probabilities, they say very little about the relative probability of different knot types.  
To address these questions we need to know about the subdominant terms.  
It is believed~\cite{orlandini98} that
\begin{equation}
  \label{eq:intro_knot_probability}
   p_N = C N^{\alpha - 3} \mu^N(1+o(1)),
\end{equation}
and it is reasonable to guess that
\begin{equation}
  \label{eq:intro_unknot_probability}
   p_N^0 = C_0 N^{\alpha_0 - 3} \mu_0^N(1+o(1)),
\end{equation}
where $\mu_0 < \mu$ and where there is numerical evidence suggesting that $\alpha_0 = \alpha$~\cite{orlandini96,baiesi2010}.  
Similarly, there is numerical evidence \cite{orlandini96,orlandini98,baiesi2010,baiesi12} that 
\begin{equation}
p_N(K) = C_K N^{\alpha_0 +n_{\mathrm{p}}(K) -3} \mu_0^N(1+o(1)),
\end{equation}
where $n_{\mathrm{p}}(K)$ is the number of prime knots in the knot decomposition of $K$.  
Thus all knot types exhibit an exponential growth rate, with the exponent depending only on the number of prime knots in the knot decomposition, and not on the particular knots involved.

The probability that a lattice polygon has knot type $K$ is (assuming that $\alpha_0 = \alpha$)

\begin{equation}
\label{eq:lattice_fraction}
{\mathrm{Prob}_N(K)} = p_N(K) / p_N = A_K N^{n_{\mathrm{p}}(K)} (\mu_0/\mu)^N(1+o(1)),
\end{equation}
where $A_K=C_K/C$, while the relative probability of the knot type being $K_1$ or $K_2$ is
\begin{eqnarray}
   {\mathrm{Prob}_N(K_1)}/{\mathrm{Prob}_N(K_2)} & = & p_N(K_1)/p_N(K_2) \nonumber \\
   &=&  (A_{K_1}/A_{K_2} )N^{n_{\mathrm{p}}(K_1)-n_{\mathrm{p}}(K_2)} (1+o(1)),
\end{eqnarray}
even if $\alpha_0 \neq \alpha$.

Our Ansatz (\ref{eq:ansatz}) for random polygons strongly resembles (\ref{eq:lattice_fraction}), has the negative exponential with $N_K =1/ \log(\mu/\mu_0)$, consistent with $N_K$ being independent of $K$.
Our form of $v_K = v_0+n_{\mathrm{p}}(K)$ with $-1/2 < v_0 < 0$ indicates that for random polygons, the analogue $\alpha > \alpha_0$.
We will give numerical evidence for this in the following.
Readers uninterested in the details of the dataset generation can skip to Section \ref{sec:results}.

\section{Methodology and datasets}\label{sec:methods}

This section describes the numerical methods used to generate closed equilateral random walks, and the knot invariants used to identify knot types.
Our numerical implementation of both the random walk models of Section~\ref{sec:random_walk_models}, and the topological invariants of Section~\ref{sec:methods_topological_identification}, are publicly available in the pyknotid knot identification toolkit~\citep{pyknotid}. 
We also perform a range of least square fits to the numerical data, using standard nonlinear fitting routines~\cite{scipy}.

\subsection{Random walk models}
\label{sec:random_walk_models}

A typical algorithm generating general random walks does not give closed loops, i.e.~curves which return to their starting point.
It is more difficult to sample the subset of closed random walks properly, but many algorithms have been proposed for generating random polygons, either equilateral or with some distribution of step lengths (such as Gaussian distribution~\citep{diao94}).
Examples include the polygonal fold, hedgehog, triangle, and crankshaft methods~\citep{alvarado11}.  
Although easy to implement numerically, not all of these algorithms give the desired probability
distribution.  When they do, they do so only as the limit distribution of a 
Markov process, and convergence may be slow ~\citep{cantarella16}. 
In particular, different algorithms appear to generate very different selections of knot types, even with parameters that are nominally similar~\citep{alvarado11}. 
For a detailed investigation of knot statistics, it is desirable to generate random polygons with a properly defined distribution.

A small number of algorithms have been shown to produce the correct distribution in polygon space rigorously. 
One method is to generate each polygon edge at random, conditioned that the walk will return to its origin after a fixed number of further steps~\citep{moore05,diao12}. 
Although good for short random knots, it is numerically complex and slow to generate longer polygons~\citep{moore05}. 
An improvement was recently proposed by~\citet{cantarella16}, in which the complicated numerical arithmetic is replaced by a direct rejection sampling of valid states, generating valid polygons with $N$ edges in $O(N^{5/2})$ time. 
This \emph{action-angle method} is the chosen source of random equilateral polygons here. 
Another approach---the `toric symplectic Markov chain Monte Carlo' algorithm \cite{cantarella16-2}---has been shown to converge to the appropriate distribution, but this is again relatively difficult to implement numerically.

We sampled $1.96 \times 10^9$ equilateral random polygons using the action-angle method, at lengths from 6 to 4000 edges. 
The sampled lengths are every $N$ from $6 \leq N \leq 50$, steps of $10$ from $50 < N \leq 200$, steps of $50$ from $200 < N \leq 1000$, and steps of $100$ from $1000 < N \leq
4000$. 
At each length $N \leq 3000$, we analysed at least $10^7$ different polygons, in some cases far more. 
For each length $N > 3000$ we analysed at least $10^6$ different polygons. 
Our analysis with the quaternionic model was based on similar choices.

\subsection{Methods for identifying knot types}
\label{sec:methods_topological_identification}

Knots abound in random walks, and it is necessary to distinguish their distinct knot types. 
The Rolfsen table of knots~\citep{rolfsen76}, with standard extensions for knots with up to 16 crossings~\citep{hoste98} denotes the knot $K_i$ as the $i$th knot with crossing number $K$, the minimum number of crossings a 2-dimensional diagram of the knot can have, which we denote $n_{\mathrm{c}}$ (see Figure~~\ref{fig:knots_and_random_walks}(a)). 
The ordering of index $i$ is effectively arbitrary.
The knot $0_1$ is the special case, called the \emph{unknot}, representing the topologically trivial, simple circle.  
Knots with a crossing number $n_{\mathrm{c}} \ge 11$ are referred to as $K_{ai}$ or $K_{ni}$ (e.g.~$11_{a343}$, $11_{n3}$), where $a$ or $n$ indicates
alternating and nonalternating knots respectively. 
Distinct chiral pairs of knots are not distinguished. 
Tables of knots and their properties are available from the Knot Atlas~\citep{knotatlas} and KnotInfo~\citep{knotinfo}. 
Knot tables give only the \emph{prime} knots, which can also be joined together by a connect sum to form \emph{composite} knots. 
Connect sums are denoted by $\#$ or with exponents denoting repeated connect sums of the same knot type. 
For instance, $3_1\#4_1^2$ represents the connect sum of a trefoil knot $3_1$ and two figure-eight knots $4_1$. 
Figure~\ref{fig:knots_and_random_walks}(a) shows the seven prime knots with $n_{\mathrm{c}} \le 6$.
Beyond these the number of knot types grows more rapidly; there are then $7$ knots with $n_{\mathrm{c}} = 7$, $21$ with $n_{\mathrm{c}} = 8$, $49$ with $n_{\mathrm{c}} = 9$, $165$ with $10$, $552$ with $11$, $2176$ with $12$, $9988$ with $13$, $46972$ with $14$, $\ldots$. 
The overall trend is of exponential growth in the number of prime knots with $n_{\mathrm{c}}$ crossings~\cite{welsh91,ernstsumners}.

Figure~\ref{fig:knots_and_random_walks}(b),(c) shows some examples of the knots $3_1$ and $7_6$ in random walks with different lengths. 
The trefoil knot $3_1$ is usually very small, made of only a few edges of the whole polygon. 
The knot $7_6$ is somewhat more complicated, dominating much of the structure of the random walk at $50$ or even $100$ edges, but as $N$ grows, the knotted regions occupy less of the curve in both cases; this behaviour for fixed knot type is well established~\cite{katritch00}. 
Furthermore, it becomes relatively unlikely that a long polygon will contain a single knot component; at large $N$, composite knots dominate the statistics, with knots occurring essentially independently in different regions of the polygon \cite{orlandini96,orlandini98}.

Determining the type of a complex random knot can be difficult.
It is most efficient to identify knots by some set of \emph{knot invariants}, i.e.~tabulated functions of knot type. 
Unfortunately, easily calculable knot invariants are not perfect discriminators, taking the same value for distinct knot types. 
Furthermore, more discriminatory invariants usually require increased computational complexity.

The most common invariant for studying random knotting is the \emph{Alexander polynomial} $\Delta_K(t)$ for knot type $K$~\citep{orlandini07,rolfsen76,adams99}. 
As numerical polynomial arithmetic is inconvenient, it is common to use the \emph{knot determinant} $|\Delta_K(-1)|$. 
Unfortunately, the knot determinant is far less discriminatory than the full Alexander polynomial: $|\Delta_{4_1}(-1)| = |\Delta_{5_1}(-1)| = 5$, whereas the simplest indistinguishable pair by Alexander polynomials is $\Delta_{6_1}(t) = \Delta_{9_{47}}(t) =t^2-5t+1$, and the simplest knot with Alexander polynomial indistinguishable from the unknot, $\Delta(t) = 1$, is $11_{n39}$. 
Therefore the determinant is often paired with certain \emph{Vassiliev invariants} $v_2, v_3, v_4, \dots$~\citep{deguchi93,deguchi94,moore04,moore05}. 
These may be calculated in polynomial time in the number of crossings of the knot representation. 
In practice, $v_2$ and the determinant are easily calculated, $v_3$ is practically calculable for knots with up to a few tens or hundreds of crossings, and higher Vassiliev invariants are generally not computationally practical for use with complicated curves. 
The Alexander polynomial is not completely independent of these invariants; in fact, $v_2$ is equal to the coefficient of $t^2$ in the (properly normalised) Conway polynomial. 
Although other invariants, such as the Jones and HOMFLY polynomials, are more powerful discriminators, computing these is exponential in the number of crossings of the projection~\citep{adams99}, and they are only practical for projected curves with no more than a few tens of crossings.

The invariants we use here are the Alexander polynomial at certain roots of unity, 
\begin{equation}
   \Delta_r \equiv \Delta_{K,r} = |\Delta_K\left(\exp(2\pi \rmi / r)\right)|, \quad r \in \mathbb{Z}.
\end{equation}
Each $\Delta_r$ is an invariant as easily calculated as the knot determinant, with the only numerical change being the use of complex datatypes. 
$\Delta_1 = 1$ always, so is not a useful invariant~\cite{burde02}, and $\Delta_2$ is the knot determinant.
As shown in \ref{appendix:roots_of_unity}, $\Delta_2,\Delta_3,\Delta_4$ conveniently are always integers, and we limit our calculation to these values. 
Higher-order roots of unity provide relatively little extra discriminatory value; to discriminate between the prime and composite knots which appear in random walks, and the first three roots of unity are almost as good as the full Alexander polynomial.
Although we could attempt to increase discriminatory power by calculating Vassiliev invariants, $v_2$ adds little to no useful discriminatory power, and $v_3$ and higher invariants significantly slow down the calculations for knots longer than a few hundred steps.

Hence, to recognise knots, we calculate $\Delta_2 = |\Delta_K(-1)|$, $\Delta_3 = |\Delta_K(\exp(2\pi \rmi/3))|$ and $\Delta_4 = |\Delta_K(\rmi)|$.
This allows us, with confidence, to distinguish all prime knots with $n_{\mathrm{c}} \le 7$, the 21 knots with $n_{\mathrm{c}} = 8$ except for $8_5, 8_{10}, 8_{11}, 8_{15}, 8_{18}, 8_{20}, 8_{21}$, and the 49 nine-crossing knots except for $9_2, 9_8, 9_{12}, 9_{16}, 9_{23}, 9_{24}, 9_{28}, 9_{29}, 9_{37}, 9_{38}, 9_{39}, 9_{40}, 9_{46}, 9_{48}$.
These excluded eight- and nine-crossing knots knots have invariants the same as either a simpler (more common) prime knot, or a common composite knot.  
We also identify composite knots with five and fewer components, involving any number of trefoil knots $3_1$ with one other prime knot, and a smaller number of examples involving more non-trefoil components. 
This introduces some error into the count e.g.~some cases identified as $3_1^2$ might be $8_{20}$ (which has the same $\Delta(t)$). 
However, in all important cases, one of the possible knots for a given set of invariants occurs with much more frequency than the alternatives, and this conflict
does not appear to harm the results.

\section{Numerical results for knot probabilities}
\label{sec:results}

\subsection{Summary of observed behaviour}

\begin{figure}
\includegraphics[width=\textwidth]{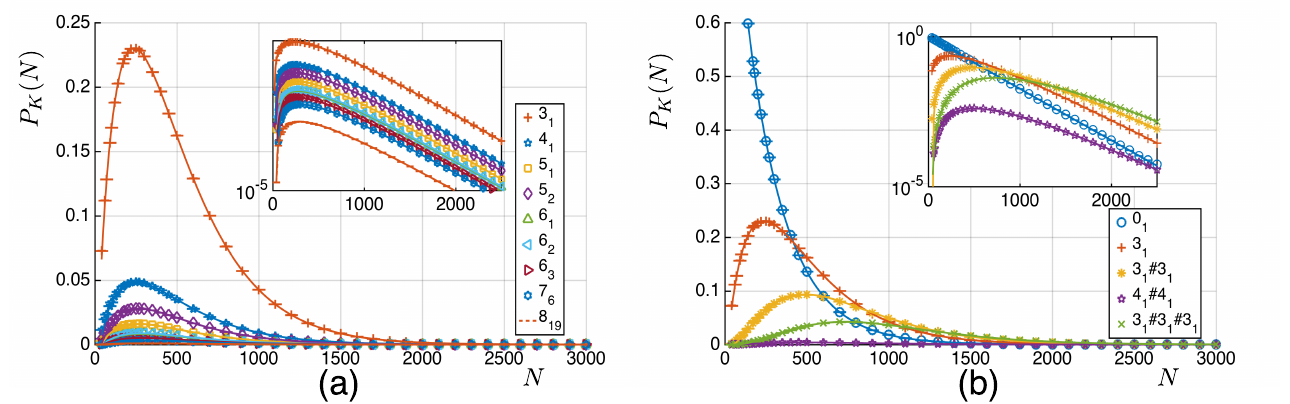}
  \caption{Fractions of different knot types in closed equilateral random walks as a function of $N$. 
  (a) Probabilities for various prime knots. 
  Note that the order by frequency is different from the ordering in the knot table. 
  The inset shows $\log P_K(N)$ against $N$, showing the similarity of behaviour for all prime knots.
  (b) Probabilities for various composite knots (including the unknot). 
  The inset shows $\log P_K(N)$ against $N$, and the near-linear slopes depend only on the number of components $n_{\mathrm{p}}(K)$.
  The plots are fitted according to (\ref{eq:ansatz}), as described in the main text.}
  \label{fig:knot_fractions}
\end{figure}

Figure~\ref{fig:knot_fractions} shows the knot fractions for several different prime and composite knot types from our numerically generated equilateral random polygons.
Figure~\ref{fig:knot_fractions} (a) shows data for prime knots.
Evidently, the prime knot probabilities are all very similar, apart from the overall amplitude factor given by $C_K$, which decreases as the knot complexity increases (as characterised by the crossing number).
Figure~\ref{fig:knot_fractions} (b) shows data for composite knots, with numbers of components $n_{\mathrm{p}}(K)$ varying from $0$ (the unknot $0_1$) to $3$ (the connect sum of three trefoils).
Composite knots with the same number of components $n_{\mathrm{p}}$, have broadly similar probabilities, up to a relative scaling determined by $C_K$.
The location of the maximum in the probability distribution increases with $n_{\mathrm{p}}$ as the overall amplitude decreases.  Overall, knots with larger $n_{\mathrm{p}}$ are less likely. 

The knot types shown in Figure~\ref{fig:knot_fractions} are only a small sampling of the data we have, and the behaviour for other knot types is consistent with that shown in the figure.
The data points in the figure are fitted according to~\eqref{eq:ansatz}, with $N_K = N_0 = 259.3\pm 0.2$, $v_K = v_0 + n_{\mathrm{p}}(K)$ with $v_0 = -0.190\pm 0.001$.
Values of $C_K$, $\beta_K$ and $\gamma_K$, are chosen to give the best fit for each knot type, and the fit for each knot type is excellent. 
The following discussion will provide more details for motivating the form of the Ansatz and the universal nature of $N_K = N_0$, $v_K = v_0 + n_{\mathrm{p}}(K)$.

Equation \eqref{eq:ansatz} is an excellent fit to the data for the various prime and composite knots shown. 
In the following sections, we will provide separate motivation to support the form of \eqref{eq:ansatz}. 
In the following subsection, we consider ratios of probabilities of knot types with the same $n_{\mathrm{p}}$, or differing by unity; the gradients being zero or one (within error) indicate the universality of $N_K = N_0$ and $v_K = v_0 + n_{\mathrm{p}}(K)$.
In the following section, we consider the best fit result for $N_0$ and $v_0$ together, showing the best fit agrees for different knot types.
We then explore this fit further for the unknot, for which the form (\ref{eq:ansatz}) with nonzero $v_0$ is new.
We then consider the different values of the amplitudes $C_K$, before discussing the corrections to scaling in the final section.

\subsection{Probability ratios}
\label{sec:ratios}

\begin{figure}
\includegraphics[width=\textwidth]{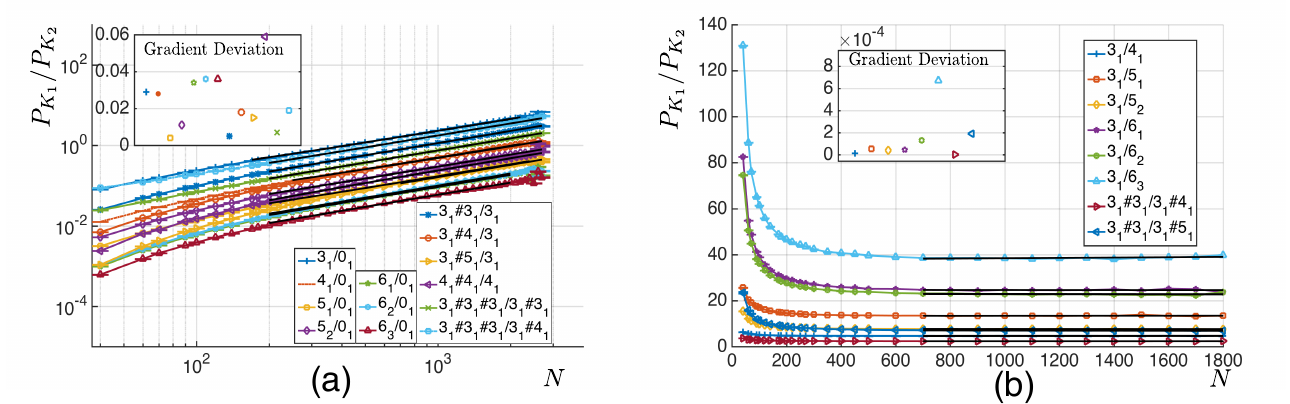} 
  \caption{
  Relative probabilities $P_{K_1}(N)/P_{K_2}(N)$ of different knot types, depending on $N$.
  (a) $P_{K_1}(N)/P_{K_2}(N)$ plotted in logarithmic scale against $N$ on a logarithmic scale for several choices of $K_1$ and $K_2$, with $n_{\mathrm{p}}(K_1) = n_{\mathrm{p}}(K_2) + 1$.
  The inset shows the modulus of the deviation from unity of the best fit line; all are clearly very close to gradient $1$. 
  The fitted lines here are from $N \approx 200$ to $N \approx 2400$.
  The errors on the fitted gradient of $1$ are $< .04$ for each, with several much better than this.
  (b) shows $P_{K_1}(N)/P_{K_2}(N)$ for several pairs where $n_{\mathrm{p}}(K_1) = n_{\mathrm{p}}(K_2)$.
  The fits here are for $700 \le N \le 1800$.
  Now the inset shows the gradient deviation from $0$, with an error $< 10^{-3}$ in each case.
  }
  \label{fig:relative_probabilities}
\end{figure}

Comparisons of $P_{K_1}(N)/P_{K_2}(N)$, for prime $K_1$ and $K_2$, justify our claim that $N_K$ and $v_K$ are independent of prime knot type.
If $N_K$ depends on knot type, then as $N \to \infty$, the ratio tends to zero or infinity exponentially rapidly.  
If $N_{K_1} = N_{K_2}$, but the exponent $v_K$ depends on prime knot type, then the ratio goes to zero or infinity, but not exponentially rapidly.  
If $N_{K_1} = N_{K_2}$ and $v_{K_1} - v_{K_2} = n_{\mathrm{p}}(K_1) - n_{\mathrm{p}}(K_2)$ then the ratio has the form
\begin{equation}
\label{eq:RatioPrime}
\frac{P_{K_1}(N)}{P_{K_2}(N)}= \frac{C_{K_1}}{C_{K_2}}N^{n_{\mathrm{p}}(K_1) - n_{\mathrm{p}}(K_2)}\left(1+ \frac{\beta_{K_1}-\beta_{K_2}} {N^{1/2}} + O(N^{-1}) \right).
\end{equation}

The results from our data for $ P_{K_1}(N)/P_{K_2}(N)$ against $N$ on a log-log scale are shown in Figure \ref{fig:relative_probabilities}.
In (a), $n_{\mathrm{p}}(K_1) = n_{\mathrm{p}}(K_2) +1$ for several pairs with $n_{\mathrm{p}}(K_2)$ = 0,1,2.
The curves fit very well to a straight line of gradient unity with a very small error in all cases.
This suggests each $K_1$ and $K_2$ have the same exponential term, and power law term differing by $1$.
In Figure \ref{fig:relative_probabilities} (b), several pairs are shown where $n_{\mathrm{p}}(K_1) = n_{\mathrm{p}}(K_2)$. 
Again the curves seem to be asymptotically linear with limiting slopes effectively zero.
None of the curves in Figure \ref{fig:relative_probabilities} (b) approach zero or infinity as $N$ increases, suggesting that $N_K = N_0$ and $v_K = v_0 + n_{\mathrm{p}}(K)$ for all knots. 
This analysis, however, does not give values for the universal constants $N_0$ and $v_0$.
The corrections to scaling $\beta_K$ in \eqref{eq:RatioPrime}, indicating the way the curve approaches the asymptotic ratios, will be considered below in Section \ref{sec:short_length_scales}.

\subsection{Determining values of $N_0$ and $v_0$}
\label{sec:fit}

\begin{figure}
  \begin{center}
  \includegraphics[width=10cm]{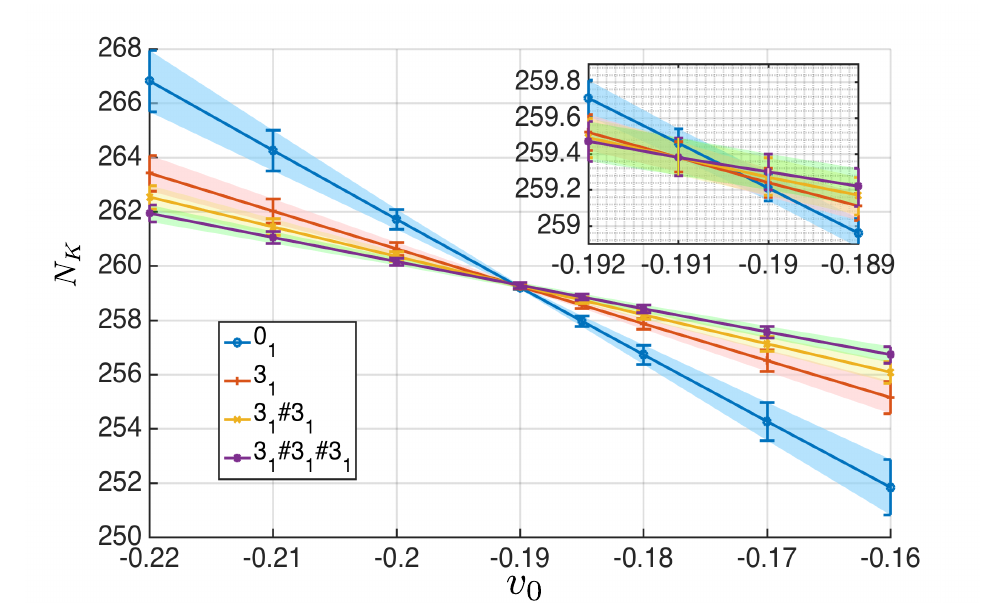}
  \end{center}
  \caption{
  Plot of fit quality for $N_K$ against $v_0 + n_{\mathrm{p}}(K)$ for the commonest knot type for each $n_{\mathrm{p}}$.
  Curves are shown for the unknot $0_1$ ($n_{\mathrm{p}} = 0$), the trefoil $3_1$ ($n_{\mathrm{p}} = 1$), and composites $3_1 \# 3_1$ ($n_{\mathrm{p}} = 2$) and $3_1 \# 3_1 \# 3_1$ ($n_{\mathrm{p}} = 3$).
  For each value of $v_0$ considered (between $-0.22$ and $-0.16$ with an increment typically of $0.01$), an error bar around the best fit $N_K$ is given.
  This is calculated based on varying $v_0$ and the corrections to scaling parameters, minimising the sum of square deviations and weighting the data points by inverse variance, within a tolerance error of $95\%$.
  The error bars for each knot type are smallest at or near the value of $v_0$ where the curves cross, detailed in the inset.
  In the inset, the lines cross at a $v_0$ between $-0.190$ and $-0.191$.
  The errors on these values are estimated based on the spread of values in the curves, with errors crossing in the inset, giving $v_0 = -0.190\pm 0.001$ and $N_K = 259.3\pm 0.2$.
  }
  \label{fig:fittingv0N0}
\end{figure}

Although justifying the general form of the knot probability, the method above does not determine the numerical values of $N_0$ and $v_0$.  
This is complicated by the fact that best fits to $N_0$ and $v_0$ cannot be determined independently.
We perform the analysis for the commonest knot type of each number of components $n_{\mathrm{p}}(K)$: the unknot $0_1$, the trefoil knot $3_1$, and connect sums of trefoils $3_1 \# 3_1$ and $3_1 \# 3_1 \# 3_1$.
As evident in Figure \ref{fig:knot_fractions}, the commonest knot types from all the data are, in order, the unknot, the trefoil and $3_1 \# 3_1$.

For each knot type and $v_0$ in the considered range, we calculate the best fit to the Ansatz (\ref{eq:ansatz}) by minimising the sum of the square deviations and weighting the data points by the inverse variance, whilst varying $N_0$, $C_K$, $\beta_K$ and $\gamma_K$.
In Figure \ref{fig:fittingv0N0} we plot the optimal $N_K$ for each $v_0$, with the error bars represent $95\%$ tolerance of the fitted data to this value (the other parameters are not shown).
The lines of best fit for $N_K$ against $v_0$ intersect very close to one another, and very close to the values where the error bars are the smallest, as shown in the inset.
The crossings do not take place at precisely the same $v_0,N_K$, and from this we estimate the errors, giving $v_0 = -0.190\pm 0.001$ and $N_K = 259.3\pm 0.2$. 

As discussed above, these values of exponents give excellent fits for all knot types, as indicated for a sample of our data in Figure \ref{fig:knot_fractions}.

\subsection{Fitting the unknot}
\label{sec:unknot}

\begin{figure}
  \begin{center}
  \includegraphics[width=10cm]{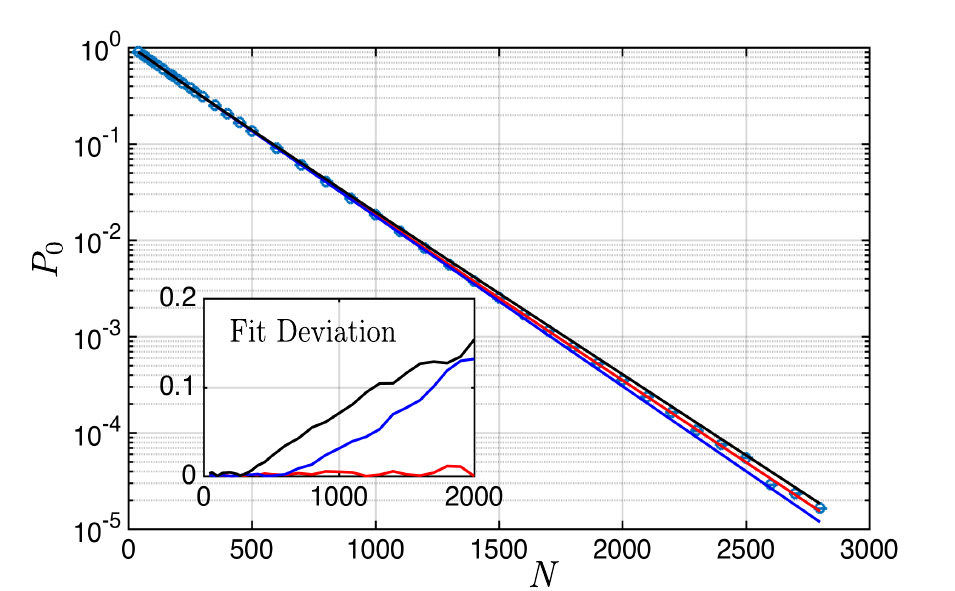}
  \end{center}
  \caption{
  Unknot fraction of closed equilateral random walks.
  Plot of $\log P_{0_1}(N)$ against $N$ for a range of $N$ up to $N = 2800$.
  The numerical data for the unknot probabilities are compared to three fits: the Ansatz (\ref{eq:ansatz}) with the numerical values fit as in Section \ref{sec:fit} (red line), the best fit assuming a pure exponential decay with the best fit exponential decay constant $N_0' = 246.5$ (blue), and the the best fit using the exponential constant $N_0 = 259.3$ (black) from Section \ref{sec:fit}.
  In each case, corrections to scaling terms are included (not shown) to optimise the fit.
  The inset shows the modulus of the relative deviation of each fit from the data.
  The Ansatz, including the power law, is a significantly better fit than the pure exponentials.
  }
  \label{fig:unknot_scaling}
\end{figure}

As we have discussed, the unknot $0_1$ appears in our Ansatz \eqref{eq:ansatz} not as a type of prime knot, but is properly considered as the unique composite knot with zero components, $n_{\mathrm{p}}(0_1) = 0$.
Without corrections to scaling, its probability is  $P_{0_1}(N) \approx C_{0_1} N^{v_0} \exp(-N/N_0)$, and indeed in the last subsection, we described how the best fit of $v_0$ and $N_0$ to the unknot data agrees very well with the values for the trefoil and its connect sums.

It has long been thought that the probability of a random unknotted polygon, ignoring small-scale effects, is simply exponential~\cite{millett94,deguchi94,millett2005,orlandini07}, 
\begin{equation}
  \label{eq:results_unknot_probability}
  P_{0_1}(N) = C_{0_1} \exp\left( - N/N_{0_1} \right)~,
\end{equation}
with no power law term.
This form is consistent with~\citep{diao95} and \eqref{eq:intro_unknot_probability}, and has been verified in a wide variety of random polygon models~\citep{deguchi94,moore04,millett2005,orlandini07,uehara17}, as well as the lattice case \cite{orlandini07,jvr90}.

In Figure \ref{fig:unknot_scaling}, we show how the unknot data for our Ansatz, including the $v_0$ exponent (and best fit corrections to scaling), compares with the raw exponential form (\ref{eq:results_unknot_probability}).
The difference between the best fits is shown in the inset. 
Most data was generated for $N \le 300$, and in this range, the agreement is good for all of the fits.
However, the pure exponential with $N_0$ systematically deviates (with a linear error) for $N > 300$, as indicated by the black curve in Figure 5, and the pure exponential with $N'_0$ deviates systematically, in a similar way, when $N > 600$, as indicated by the blue curve in Figure 5. 
The fitting exponents vary from the data systematically with different signs. 
The two fitted curves without the power law term have a systematic deviation that grows as $N$ increases, while our Ansatz, indicated by the red curve in Figure 5, is a substantially better bet without any systematic deviation.

The Ansatz with exponential and power law found from the last section, based on the data from the knots as well as unknots, gives a good fit over the entire range. 
This suggests that the pure exponential model is an approximation for small $N$, while for large $N$ it is necessary to have the power law term. 
Meanwhile, our Ansatz suggest a greater universality that incorporates the unknot into a wider class, as a composite knot with zero components.
As discussed in Section \ref{sec:fit}, these values were chosen from the simultaneous optimisation both of the unknot, and multiple trefoil knots.

In fact, the deviation from pure exponential scaling for the unknot has been observed in previous studies \cite{moore04,uehara17}.
However, its effect was not distinguished from systematic errors: many older studies do not sample enough random walks to detect the change. 
The discrepancy was interpreted~\citep{moore04} as nontrivial knots being incorrectly identified as the unknot.
These misidentifications are not represented in the error bars as it is difficult to estimate their number.

It is very difficult to estimate confidently, beyond tabulations, the number of nontrivial knots with Alexander polynomial (at roots of unity) corresponding to the unknot.
The prime knots with $\Delta_2,\Delta_3,\Delta_4$ matching the unknot with $n_{\mathrm{c}} \le 15$, grow quickly with $n_{\mathrm{c}}\ge 11$ (there are 2 examples with $n_{\mathrm{c}} = 11$, 2 with $n_{\mathrm{c}} = 12$, 15 with $13$, 36 with 14, 145 with 15).
The probability of each of these knot types occurring drops rapidly with $n_{\mathrm{c}}$, as discussed in Section~\ref{sec:Ck}; it is not clear how this decrease compares to the exponential increase of knot types with $n_{\mathrm{c}}$, and no stable pattern emerges for $n_{\mathrm{c}} \le 15$.
Composite knots consisting of these components would also appear as the unknot, but are even rarer.

We do not believe that the deviation in Figure~\ref{fig:unknot_scaling} can be explained by misidentification of unknots.
Rather than estimate the misidentification rate by improving the discriminating power, we adopt the opposite methodology, by comparing the results with a less discriminatory analysis using only the determinant $\Delta_2$ which is a much weaker invariant than the set $\Delta_2,\Delta_3,\Delta_4$.
There are 2 examples misidentified as the unknot with $n_{\mathrm{c}} = 10$, 4 with $n_{\mathrm{c}} = 11$, 11 with $n_{\mathrm{c}} = 12$, 44 with 13, 162 with 14, 724 with 15, $\ldots$. 
If the deviation from the fit in Figure~\ref{fig:unknot_scaling} were due to misidentification, we would expect a significantly larger deviation from identifying the unknot only by determinant.
However, the change to the results is very small: for instance, it accounts for $< 2\%$ more detected `unknots' at length $2000$ than with $\Delta_2,\Delta_3,\Delta_4$, and this misidentification rate grows only slowly with $N$.
This is far smaller than the $\sim 13\%$ deviation of the unknot fraction from exponential decay in Figure~\ref{fig:unknot_scaling}, despite the unknot misidentification rate being far larger than with the original data.
These are too small to be visible in any of the plots of Figure \ref{fig:unknot_scaling}, and we conclude that unknot misidentification is a negligible error in Figure~\ref{fig:unknot_scaling}.
This also suggests that the determinant alone is a reasonably reliable invariant for detecting unknotted random polygons; however, it cannot distinguish between simple prime knots.

Furthermore, our results about the unknot do not exist independently of other knot types: the argument in Section \ref{sec:ratios}.
If the unknot indeed had the different form (\ref{eq:results_unknot_probability}) with different exponents, it would be very surprising that the inclusion of other knots with the same Alexander polynomial would exactly cancel to give the relevant ratio plots in Figure \ref{fig:relative_probabilities} (b).
As discussed above, these plots of the ratio of the logarithm of probabilities of various prime knots against the unknot tending to straight lines of gradient unity, and not to $0$ or $\infty$, indicates that the unknot probability has the same form as prime knots, except for the different $n_{\mathrm{p}}$.

It is important to note that the best fits reported here involve varying the corrections to scaling parameters $\beta_K$ and $\gamma_K$, not shown in Figure~\ref{fig:unknot_scaling}.
These parameters were optimised for all three fits to the data shown.
Varying the values of $\beta_K$ and $\gamma_K$, does not affect the systematic advantage of the Ansatz fit over the others; a variation of ~$10\%$ in these fitting parameters changes the deviation by about $3\%$ at $N = 2000$.
These fitting parameters will be discussed in general later in Section \ref{sec:short_length_scales}.

\subsection{The knot coefficient amplitudes $C_K$}
\label{sec:Ck}

\begin{figure}
  \includegraphics[width=\textwidth]{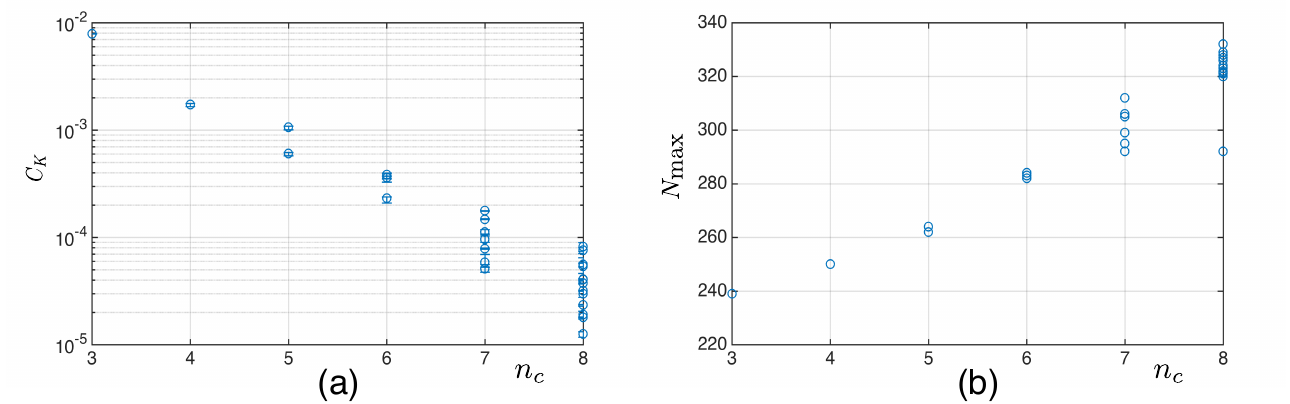}
  \caption{Features that vary with knot type.
  (a) Scatter plot of $C_K$ for the prime knots, following data given in Table \ref{table:CKp}.
  $C_K$ tends to decrease with crossing number $n_{\mathrm{c}}$, but with a broad range of $C_K$ for a given $n_{\mathrm{c}}$.
  (b) Scatter plot of $N_{\mathrm{max}}$, the position of maximum of $P(N)_K$ for different $K$.
  Without the corrections to scaling, these would all be at the same point at $N \approx 210$.
  }
  \label{fig:properties_with_K}
\end{figure}

\begin{table}[ht!]\begin{centering}
\begin{tabular}{|c||c|c|c|c|c|c|c|}
\hline
$K$ & $3_1$ & $4_1$ & $5_1$ & $5_2$ & $6_1$ & $6_2$ & $6_3$ \\
$-\log C_K$ & $4.84$ & $6.36$ & $7.41$ & $6.83$ & $7.93$ & $7.87$ & $8.37$ \\
error & $0.003$ & $0.01$ & $0.015$ & $0.015$ & $0.024$ & $0.024$ & $0.022$ \\
\hline
\hline
$K$ & $7_1$ & $7_2$ & $7_3$ & $7_4$ & $7_5$ & $7_6$ & $7_7$ \\
$-\log C_K$ & $9.88$ & $9.10$ & $9.25$ & $9.74$ & $8.82$ & $8.64$ & $9.45$ \\
error & $0.03$ & $0.03$ & $0.023$ & $0.03$ & $0.029$ & $0.024$ & $0.027$ \\
\hline
\hline
$K$ & $8_1$ & $8_3$ & $8_4$ & $8_6$ & $8_7$ & $8_8$ & $8_9$ \\
$-\log C_K$ & $10.19$ & $10.87$ & $10.36$ & $9.79$ & $10.11$ & $9.49$ & $10.66$ \\
error & $0.38$ & $0.4$ & $0.38$ & $0.2$ & $0.24$ & $0.12$ & $0.22$  \\
\hline
\hline
$K$ & $8_{12}$ & $8_{13}$ & $8_{14}$ & $8_{16}$ & $8_{17}$ & $8_{19}$ & $\phantom{x}$ \cr
$-\log C_K$ & $10.42$ & $10.11$ & $9.41$ & $10.93$ & $11.28$ & $9.83$ & $\phantom{y}$ \cr
error & $0.37$ & $0.25$ & $0.13$ & $0.4$ & $0.44$ & $0.21$ & $\phantom{z}$ \cr
\hline
\hline
$K$ & $3_1 \# 3_1$ & $3_1 \# 4_1$ & $3_1 \# 5_1$ & $4_1 \# 4_1$ & $4_1 \# 5_1$ & $3_1 \# 3_1 \# 3_1$ & $3_1 \# 3_1 \# 3_1 \# 3_1$  \\
$-\log C_K$ & $11.65$ & $12.53$ & $13.72$ & $14.77$ & $15.15$ & $18.84$ & $26.37$  \\
\text{error} & $0.07$ & $0.03$ & $0.037$ & $0.06$ & $0.05$ & 0.02 & 0.03 \cr
\hline
\end{tabular}

  \caption{
  Values of $-\log C_K$, the logarithm of knot coefficient/amplitude for random polygons, depending on knot type $K$, for the simplest distinguishable prime and composite knots.
  The unknot has $C_{0_1} = 3.67$, i.e.~ $\log C_{0_1} = 1.30$.
  These are found from the best fits for each knot type using (\ref{eq:ansatz}) with the fixed values of $N_0$ and $v_0$, and varying $\beta_K$ and $\gamma_K$ for the best fit. 
  The knot $8_{2}$ is absent from the table because since its occurrence in the data is nearly negligible, and hence a fit is not possible.
    }
  \label{table:CKp}
  \end{centering}
\end{table}

The results discussed so far indicate that the main way the knot type determines the random polygon probability is the \emph{knot coefficient} $C_K$ (up to the number of prime components $n_{\mathrm{p}}(K)$ in $K$, and ignoring corrections to scaling).
In particular, the relative fractions of composite knots with the same $n_{\mathrm{p}}$ are determined almost entirely by the \emph{knot coefficient} $C_K$~\cite{uehara17}.
The values of $C_K$ endow the (prime) knots with a natural ordering---lower values indicate more complex knots, occurring more rarely---although little is known about how $C_K$ is related to the average geometry of the curves.

We estimate the amplitudes $C_K$ for all prime knots with at most seven crossings, and some eight crossing knots and composite knots.  
The values of $C_K$ for the simplest prime and composite $K$ for random polygons are given in Table \ref{table:CKp}.
Figure \ref{fig:properties_with_K} (a), shows how the the amplitudes $C_K$ depend on crossing number $n_{\mathrm{c}}$ for prime knots. 
There is a general decrease in the value of the amplitude as the crossing number increases but the spread in values at fixed crossing number also increases as the number of prime knots with that crossing number increases.
This is also consistent with our results for the quaternionic model of random knots with varying edge lengths, given in Table \ref{table:CKq}.

\begin{table}[ht!]\begin{centering}
\begin{tabular}{|c||c|c|c|c|c|c|c|}
\hline
$K$ & $3_1$ & $4_1$ & $5_1$ & $5_2$ & $6_1$ & $6_2$ & $6_3$ \\
$-\log C_K$ & $5.14$ & $6.72$ & $7.77$ & $7.22$ & $8.30$ & $8.31$ & $8.80$ \\
error & $0.01$ & $0.02$ & $0.026$ & $0.024$ & $0.034$ & $0.036$ & $0.05$ \\
\hline
\hline
$K$ & $7_1$ & $7_2$ & $7_3$ & $7_4$ & $7_5$ & $7_6$ & $7_7$ \\
$-\log C_K$ & $10.29$ & $9.48$ & $9.60$ & $10.29$ & $9.48$ & $9.59$ & $9.85$ \\
error & $0.09$ & $0.046$ & $0.048$ & $0.084$ & $0.046$ & $0.046$ & $0.062$ \\ 
\hline
\hline
$K$ & $8_6$ & $8_{19}$ & $3_1 \# 3_1$ & $3_1 \# 4_1$ & $4_1 \# 4_1$ & $3_1 \# 3_1 \# 3_1$ &  \\
$-\log C_K$ & $10.28$ & $10.29$ & $12.47$ & $13.24$ & $15.55$ & $20.10$ & \phantom{x} \\
error & $0.079$ & $0.08$ & $0.1$ & $0.08$ & $0.06$ & $0.2$ & \phantom{y}  
\cr
\hline

\end{tabular}

  \caption{
  Values of $-\log C_K$, the logarithm of knot coefficient/amplitude for quaternionic random walks, depending on knot type $K$, for the simplest distinguishable prime and composite knots.
  The data here are less good than the random polygons.
  The unknot has $C_{0_1} = 4.25$, i.e.~ $\log C_{0_1} = 1.46$.
  These are found from the best fits for each knot type using (\ref{eq:ansatz}) with the fixed values of $N_0$ and $v_0$, and varying $\beta_K$ and $\gamma_K$ for the best fit.
    }
  \label{table:CKq}
  \end{centering}
\end{table}

It is interesting to examine the values of the amplitude ratios and we have estimated $C_{3_1}/C_K$ for prime knots $K$.  
These ratios are the relative probabilities of the two knots in the large $N$ limit.  
We find that $C_{3_1}/C_{4_1} \approx 4.6$, $C_{3_1}/C_{5_1} \approx 13.0$, $C_{3_1}/C_{5_2} \approx 7.3$, $C_{3_1}/C_{6_1} \approx 22.0$, $C_{3_1}/C_{6_2} \approx 20.7$ and $C_{3_1}/C_{6_3} \approx 34.1$.  
This suggests a trefoil is about 4.6 times more likely than a figure-eight knot in the large $N$ limit, and so on.  
These values are very different from the values found by Janse van Rensburg and Rechnitzer \cite{Rensburg-Rechnitzer} for lattice knots but they are close to the values that we find for the quaternionic knots model, for which $C_{3_1}/C_{4_1} \approx 4.85$ is fairly close to the value found by Deguchi \cite{deguchi97}.  
Similarly, for $C_{3_1\#3_1}/C_{3_1\#4_1}$ we find a value of about 2.4 and Deguchi \cite{deguchi97} reports a value of about 2.5.   
It seems that amplitude ratios are probably universal among the off-lattice models (with no excluded volume term) but that lattice knots belong to a different universality class \cite{Rensburg-Rechnitzer}.

In Figure \ref{fig:properties_with_K} we plot the values of $N$ for $N_{\mathrm{max}}$,  the maximum of $P_K(N)$ for different prime knots.
Without corrections to scaling, all of these would be at $N = N_0 (v_0+1) \approx  210$.
Evidently, the values of $N_{\mathrm{max}}$ are all larger than $210$, and increase with $n_{\mathrm{c}}$.
This shows the significant effect of the corrections to scaling terms, to which we now turn.

\subsection{Corrections to scaling}
\label{sec:short_length_scales}
\begin{figure}
\includegraphics[width=\textwidth]{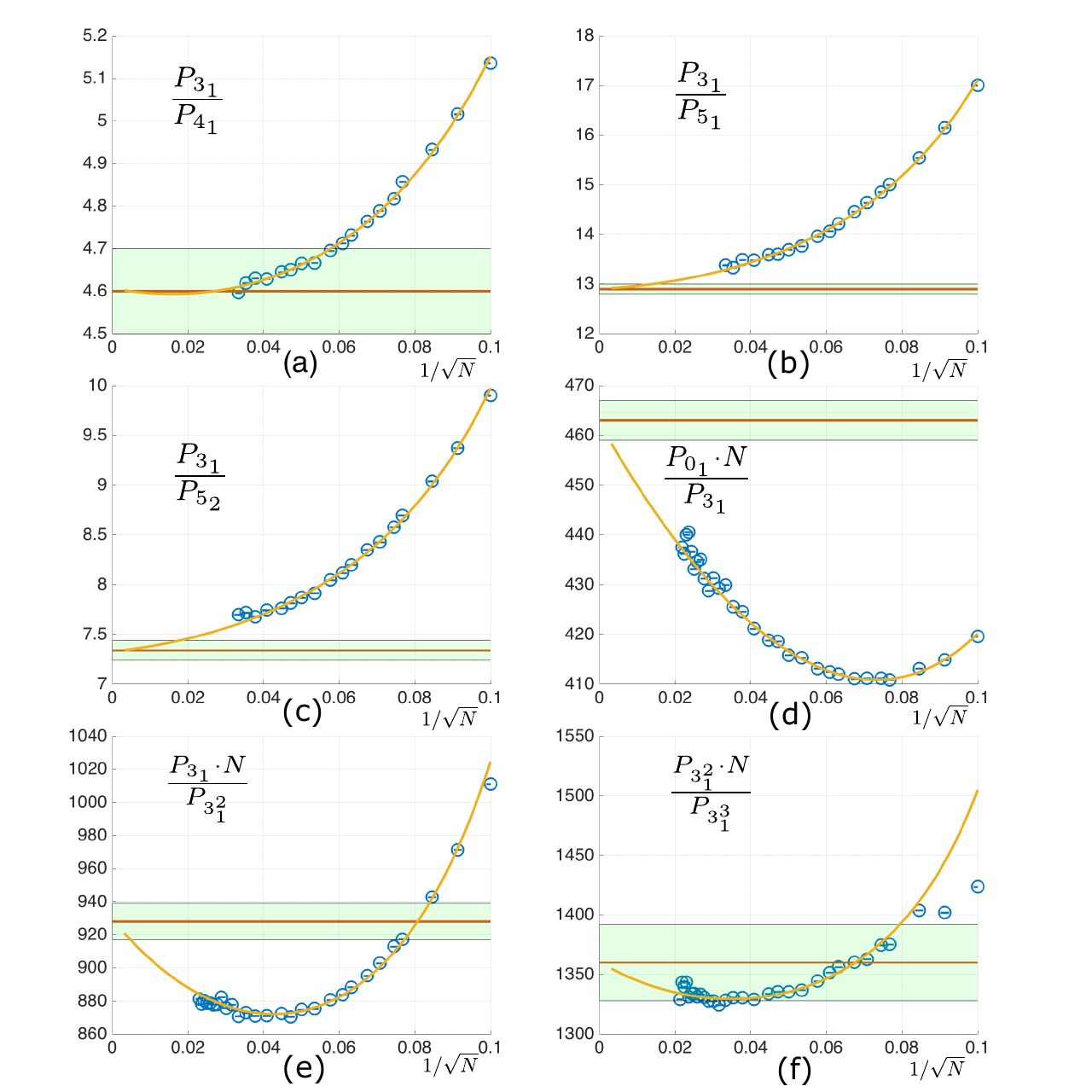} 
  \caption{Approaching asymptotic values. 
  Ratios $P_{K_1}(N) N^{n_{\mathrm{p}}(K_2)-n_{\mathrm{p}}(K_1)}/P_{K_2}(N)$ are plotted against  $1/\sqrt{N}$.
  (a)-(c), $K_1$ and $K_2$ are both prime, whereas in (d)-(f), $n_{\mathrm{p}}(K_2)-n_{\mathrm{p}}(K_1) = 1$.
  The orange curves are the best fits from the Ansatz, including the corrections to scaling.
  The constant lines (red) indicate the ratio of knot coefficients $C_{K_1}/C_{K_2}$, with errors given in the green band.
  The fact that the curves appear to approach these constant values in the limit $1/\sqrt{N} \to 0$ indicates once again that the overall scalings follow the Ansatz (\ref{eq:ansatz}) with universal $N_0$ and $v_0$.
  The approach indicates how the correction to scaling term $\beta_K$ compares between $K_1$ and $K_2$.
  In (a)-(c), which are comparisons of simple prime knots, this is close to zero, but in (d)-(f), for knots with different numbers of components, this is quite different.
  }
  \label{fig:corrections}
\end{figure}

In Ansatz (\ref{eq:ansatz}), we include corrections to scaling terms.
There are two: one, proportional to $1/\sqrt{N}$, with parameter $\beta_K$, and a Darboux-type $1/N$ with parameter $\gamma_K$.
These are suggested by the corrections expected for self-avoiding walks on a lattice.  
The coefficient $\gamma_K$ should depend on knot type since it will reflect, in part, the minimum number of edges required to tie the knot.  
It is not \emph{a priori} obvious whether $\beta_K$ should depend on knot type.

For prime knots we have presented evidence that the exponential growth term $N_K = N_0$ and the exponent $v_K = v_0 + 1$, are independent of prime knot type.
Thus the ratio of probabilities of two prime knots, as $N \to \infty$, should approach the ratio of their amplitudes.  
The corrections to scaling terms control how this limit is approached.  
In Figure \ref{fig:corrections} (a)-(c), we show the ratios of probabilities for various pairs of prime knots, as a function of $1/\sqrt{N}$.  
These curves appear to intercept the vertical axis (in the limit $N \to \infty$) at positive finite values, consistent with $N_K$ and $v_K$ indeed being independent of prime knot type, with a limiting slope close to zero.
This intercept is close to the ratio of knot coefficients $C_{K_1}/C_{K_2}$, given by the horizontal red lines.
If the limiting slope is exactly zero, then $\beta_K$ is independent of prime knot type. 

While the asymptotic curve in Figure \ref{fig:corrections} (a) is nearly flat, the asymptotic gradients in Figure \ref{fig:corrections} (b) and (c) are clearly non-zero.
This suggests that, although that the $\beta_K$ values of $3_1$ and $4_1$ are close, $\beta_K$ for $5_1$ and $5_2$ differs from that of $3_1$ much more than $4_1$. 
We infer that $\beta_K$ is dependent on knot type, and is somehow dependent on crossing number.
This result is also consistent with Figure \ref{fig:properties_with_K} (b) where the position of maximum of $P(N)_K$ for prime knots shifts to higher N as the number of crossings increases; such a shift is algebraically reflected by the correction terms.

In our fitting calculations as described in Section \ref{sec:fit}, both $\beta_K$ and $\gamma_K$ were allowed to vary in order to achieve the best fit for $N_0$.
These best fits are also shown in Figure \ref{fig:corrections}.
The best fit $\beta_K$ and $\gamma_K$ for several knot types are given in Table \ref{table:bgK}.

\begin{table}[ht!]\begin{centering}
\begin{tabular}{|c||c|c|c|c|c|c|c|}
\hline
$K$ & $3_1$ & $4_1$ & $5_1$ & $5_2$ & $6_1$ & $6_2$ & $6_3$ \\
$-\beta_K  $ & 1.24 & 1.14 & 1.30 & 1.44 & 3.63 & 3.47 & 3.79 \\
$-\gamma_K $ & 12.5 & 21.9 & 29.7 & 29.9 & 22.1 & 22.9 & 20.8 \\

\hline
\hline
$K$ & $7_1$ & $7_2$ & $7_3$ & $7_4$ & $7_5$ & $7_6$ & $7_7$ \\
$-\beta_K $ & 5.65 & 5.91 & 5.60 & 4.24 & 6.55 & 6.78 & 6.38 \\
$-\gamma_K$ & 9.97 & 10.3 & 12.0 & 22.5 & 5.40 & 4.03 & 7.43 \\
\hline
\hline
$K$ & $8_1$ & $8_3$ & $8_4$ & $8_6$ & $8_7$ & $8_8$ & $8_9$ \\
$-\beta_K $ & 8.77 & 10.24 & 8.65 & 9.18 & 8.68 & 9.66 & 8.92 \\
$\gamma_K$ & 8.67 & 20.6 & 7.96 & 10.8 & 7.26 & 15.28 & 9.50 \\ 
\hline
\hline
$K$ & $8_{12}$ & $8_{13}$ & $8_{14}$ & $8_{16}$ & $8_{17}$ & $8_{19}$ &  \\
$-\beta_K $ & 9.78 & 9.15 & 10.16 & 8.31 & 9.62 & 7.10 & \\
$\gamma_K$ & 16.2 & 12.2 & 18.8 & 2.64 & 15.2 & 5.7 & \\ 
\hline

\end{tabular}

  \caption{
  Values of correction to scaling coefficients $\beta_K$ and $\gamma_K$ for random polygons when $K$ is prime.
  Optimising these values was part of the fitting procedure.
  In addition, the unknot is found to have values $\beta_{0_1} = -3.8$, $\gamma_{0_1} = 8.3$.
  Furthermore, simple composite trefoil knots have $\beta_{3_1^2} = +1.7$, $\gamma_{3_1^2} = -48.9$ and $\beta_{3_1^3} = +3.3$, $\gamma_{3_1^3} = -69$.
  We find the analogous terms for the quaternionic walks to follow similar trends. 
    The knot $8_{2}$ is absent from the table because since its occurrence in the data is nearly negligible, and hence a fit is not possible.
    }
  \label{table:bgK}
  \end{centering}
\end{table}

The best fit values for $\beta_K$ are given in the table to be $-1.24, -1.14$ and $-1.3$ for $3_1$, $4_1$ and $5_1$, which are very close but not quite the same; this is consistent with the fitted curves in Figure \ref{fig:corrections} (a)-(c) having a small gradient when they meet the vertical axis.

We also compare $N P_{K_1}(N)/ P_{K_2}(N)$ against $1/\sqrt{N}$ where $n_{\mathrm{c}}(K_2)-n_{\mathrm{c}}(K_1) = 1$. 
These are shown in  Figure \ref{fig:corrections} (d)-(f).
The curves approach a positive finite value as $1/\sqrt{N} \to 0$, consistent with our claims that $N_{K_1} = N_{K_2}$ and $v_{K_2} = v_{K_1} + 1$.  

The unknot best fit of $\beta_{0_1} = -3.8$, which is quite far from $\beta_{3_1}$, consistent with the gradient in Figure \ref{fig:corrections} (d).
When we compare the unknot against the trefoil, or composites of trefoils against each other, there is a strong negative gradient, suggesting the $\beta_{K_2} > \beta_{K_1}$.
The strong minimum in each case indicates a significant effect from the Darboux term $\gamma_K$ as well.

We also found (not shown) that when $K_1$ and $K_2$ are composite with the same number of components, the $\beta_K$ values are similar when their components have similar $\beta_K$ values.

\section{Summary of results from quaternionic random walks}
\label{sec:quat}

\begin{figure}
  \includegraphics[width=\textwidth]{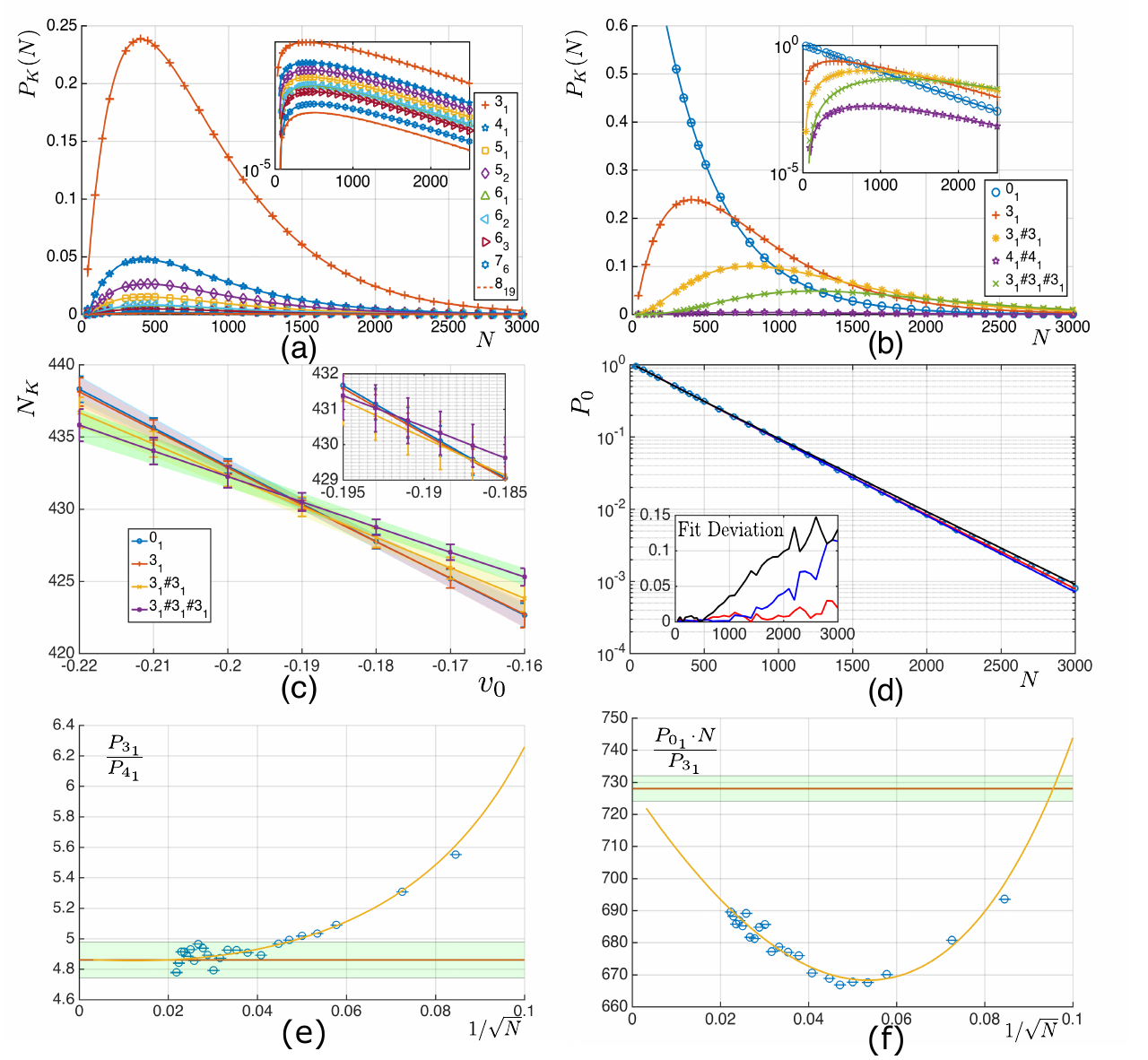}
  \caption{Summary of results for closed quaternionic random walks, analogous to previous figures for equilateral random polygons.
  (a) Probabilities $P_K(N)$ plotted against the best fit of the Ansatz for various prime knots, and (b) for various composite knots.
  (c) Best fit of $N_0$ and $v_0$.
  The optimal value of $N_0$ is different from random polygons, but is similar for the power law exponent $v_0$.
  (d) Comparison of unknot probability $P_0(N)$ analogous to Figure \ref{fig:unknot_scaling}. 
  (e) Plot of ratio of probability of trefoils to figure-8 knots against $1/\sqrt{N}$.
  The curve approaches the asymptotic value with a small gradient, consistent with the $\beta_K$ terms having similar values.
  (f) Plot of ratio of probability of unknots times $N$ to trefoils, against $1/\sqrt{N}$.
  As for the analogous polygon data in Figure \ref{fig:corrections} (d), the $\beta_K$ values are clearly different.
  }
  \label{fig:quaternionic}
\end{figure}

In Figure \ref{fig:quaternionic} we summarise the main results of the analysis with quaternionic random walks.
These all appear very similar to the analogous plots for equilateral random polygons; the fits from the Ansatz in Figure \ref{fig:quaternionic} (a), (b) look very good, and the best fit for $v_0$ is at $-0.19$, albeit with a slightly less good fit than for equilateral random polygons.
The optimal value of the exponential decay constant $N_0 = 430.5$, somewhat larger than for equilateral random polygons (as expected for walks with varying step lengths, since multiple short steps do not contribute significantly to the knotting topology).

Figure \ref{fig:quaternionic} (d) shows that the best fit line for the unknot probability $P_0(N)$ again follows our Ansatz (red) much better, for larger $N$, than either the pure exponential with the value of $N_0$ fitted from Figure \ref{fig:quaternionic} (c) (black) or the best fit pure exponential (blue).

Figure \ref{fig:quaternionic} (e) shows the ratio of probability of trefoils to figure-8 knots decreases to the ratio of knot coefficients, apparently with similar values of $\beta_K$.
Just as for equilateral random polygons, the $\beta_K$ are quite different for prime knots and the unknot, evident in Figure \ref{fig:quaternionic} (e).
Again, the lower quality of the large-$N$ data for quaternionic walks is revealed towards the asymptotic regimes.

We give values of the best fit $C_K$ values in Table \ref{table:CKq}, which were considered briefly in Section \ref{sec:Ck}.

\section{Discussion}
\label{sec:discussion}

We have presented numerical evidence that Equation (\ref{eq:ansatz}) describes the scaling behaviour of the probability of different knot types occurring in random polygons with length $N$, that is, an exponential decay characterised by constant $N_0$ (the same for all knots, but dependent on the model), and a power law term $v_K = v_0 + n_{\mathrm{p}}(K)$ depending on the number of prime components $n_{\mathrm{p}}(K)$ of $K$, but with $v_0 \approx -0.190$ a universal constant.
They also depend on a knot coefficient/amplitude $C_K$, depending on knot type, and terms giving correction to scaling for smaller $N$.
These results are inspired by and are similar to the corresponding results for knots in random polygons on lattices \cite{orlandini96,orlandini98}.

In particular, we have provided firm evidence of $v_0$ providing a power law correction for the unknot scaling (consistent with the unknot being the unique knot with $n_{\mathrm{p}} = 0$).
The lattice result, and evidence from previous numerical surveys (over a smaller range of $N$), gave the unknot probability as a pure exponential.

Our investigations highlight two contrasting types of result. 
Firstly, our unusual numerical precision has led to surprising new observations about knotting of random polygons; in particular the probability of unknotting does not simply decay exponentially with side length $N$, in contrast with many other studies. 
Secondly, this numerical accuracy reveals fundamental limitations of this type of knotting analysis.
Only knots with minimum crossing number $n_{\mathrm{c}} \le 16$ are classified, yet the number of distinct possible knot types a polygon can assume, grows very rapidly with $N$.
Future advances in numerical resources are unlikely to extend to dramatically longer lengths, without accompanying advances in knot recognition. 
As discussed previously, the main results here are given for the action-angle model of equilateral random polygons, although our preliminary data for quaternionic random walks (which were almost as extensive) qualitatively support all the numerical observations and results.

The largest systematic error in this kind of analysis is knot misidentification.
In our investigation of unknot probability, we found that the unknot misidentification rate appears to be almost irrelevant, and is no larger than the other errors. 
The occurrence of knots with the same Alexander polynomial (or invariants $\Delta_2, \Delta_3, \Delta_4$) as another with equal or lower $n_{\mathrm{c}}$ seems to become an effect around $n_{\mathrm{c}} = 11$; the number of knots begins to grow quickly here (552 prime knot types, compared to 165 at $n_{\mathrm{c}} = 10$). 
It isn't clear how the misidentification rate could be improved, as more powerful knot invariants such as the Jones polynomial are especially slow to calculate for the long curves that present most of the problems. 
The third-order Vassiliev invariant $v_3$ (and possibly others of higher order) is at least a polynomial time invariant, but as the polynomial order is higher, these are still relatively slow to calculate. 
It is also possible that new, polynomial time invariants might provide extra discriminating power, such as the new example introduced in~\cite{bar-natan17}, but it is not yet clear what the improvements in knot resolution could be with these.

The results we report are strongly backed by numerical evidence, and hopefully will stimulate new investigations into proving them rigorously.
Following the results for lattices, our results are consistent with 
\begin{itemize}
\item the existence of a pattern theorem for unknotted equilateral polygons;
\item the tightness of individual prime components;
\item different prime components occurring almost independently along the polygon.
\end{itemize}

In spite of this, the meaning of $C_K$ for prime knots remains largely mysterious.
It is impossible for \emph{all} random polygon knots to have a maximum probability at $N \approx 210$ -- knots of a sufficiently large crossing number will not be possible in a polygon with 210 sides.
The correction to scaling $\beta_K$ and $\gamma_K$ are not fundamental, but are indicative of various asymptotic terms beyond leading order about which we have no knowledge.
The curious asymptotic behaviour in Figure \ref{fig:corrections} emphasises unusual trends in the largest $N$ data in which we have confidence.
Numerical resolution of all such questions is clearly beyond our current capabilities.

\ack
We are grateful to Jason Cantarella, Tetsuo Deguchi, David Foster, Enzo Orlandini and Eric Rawdon for discussions, and to Keith Alexander for providing the knot diagrams in Figure \ref{fig:knots_and_random_walks} (a).
This research was funded in part by the Leverhulme Trust Research Programme Grant No. RP2013-K-009, SPOCK: Scientific Properties of Complex Knots.

\vspace{.3cm}
\noindent The datasets generated and analysed in the paper are available from the authors.

\appendix

\section{The Alexander polynomial evaluated at roots of unity}
\label{appendix:roots_of_unity}

As discussed in Section~\ref{sec:methods_topological_identification}, the Alexander polynomial $\Delta_K(t)$ of a knot $K$ can be evaluated at roots of unity $t = \exp(2 \pi \rmi / r)$ for integer $r$. 
In particular, we claim that when $r=2,3,4$ then $\Delta_K(\exp(2 \pi \rmi / r)$ is always an integer, as proved as follows.

The coefficient sequence of any Alexander polynomial $\Delta_K(t)$ is palindromic around a single central term~\cite{rolfsen76,adams99,livingston93}: for instance, 
$\Delta_{3_1}(t) = 1 - t + t^2$, and $\Delta_{6_3}(t) = 1 - 3t + 5t^2 - 3t^3 + 1$. 
Since the overall order of $\Delta_K(t)$ is arbitrary, any Alexander polynomial can therefore be rearranged to the form
\begin{eqnarray}
  \Delta_K(t) &= & a t^n + b(t^{n-1} + t^{n+1}) + c(t^{n-2} + t^{n+2}) +   \ldots \\
  &= & t^n\left(a + b (t^{-1} + t) + c (t^{-2} + t^2) + \ldots\right)~.
\end{eqnarray}
The overall factor $t^n$ is irrelevant to the invariants $\Delta_r = |\Delta_K(\exp(2 \pi \rmi / r)|$.

Each coefficient except $a$ is multiplied by $t^{-p} + t^p = 2 \cos(2\pi p / r)$ for some integer $p$. 
This is clearly an integer when $r = 2,3,4$ (for which $\cos(2\pi / r)$ = $-1,-\frac{1}{2},0$ respectively).
However, these are special values of $r$ and in general $|\Delta_K(\exp(2 \pi \rmi / r))|$ is not an integer.

The Alexander polynomial evaluated at each root of unity can be  considered as the sum of coefficients with a particular weighting: for instance, the
determinant $\Delta_2$ when $r=2$ is the sum of the coefficients with alternating sign. 
Other values give more complicated sequences.



\end{document}